%% file: arxiv_v2.tex
\documentclass[runningheads]{llncs}

\usepackage[T1]{fontenc}
\usepackage{graphicx}

\usepackage{amsmath, amssymb, mathtools}
\DeclareMathOperator*{\argmax}{\textnormal{arg\,max}}

\usepackage{algorithm}
\usepackage{algorithmic}

\usepackage{booktabs}

\usepackage{enumitem}

\usepackage{xcolor}
\usepackage{tikz}
\usetikzlibrary{shapes,patterns,calc,spy,arrows,automata,math,backgrounds,
decorations.pathreplacing,decorations.markings,arrows.meta,positioning}
\usepackage{tikz-network}
\usepackage{pgfplots}
\pgfplotsset{compat=1.18}

\usepackage{fontawesome5}
\usepackage{tcolorbox}
\tcbuselibrary{listings,breakable}

\newcommand{\xorproxy}{DNF learning proxy}
\newcommand{\omegaXOR}{\omega_{\textsc{xor}}}
\newcommand{\omegaVDone}{\omega_{\textsc{vd1}}}
\newcommand{\omegaVDtwo}{\omega_{\textsc{vd2}}}
\newcommand{\omegaNVD}{\omega_{\textsc{nvd}}}
\newcommand{\omegaH}{\omega_{\textsc{h}}}

\newtcblisting{MyVerbatim}{
breakable,
colframe=gray!75,
colback=white,
listing only,
fontupper=\small\ttfamily,
left=5pt,
right=5pt,
top=5pt,
bottom=5pt,
before skip=10pt,
after skip=10pt,
listing options={
  basicstyle=\small\ttfamily,
  breaklines=true,
  tabsize=4,
  numbers=none,
  numbersep=5pt,
  showspaces=false,
  showtabs=false,
  frame=none,
  keywordstyle=\color{black},
  commentstyle=\color{gray},
  stringstyle=\color{black},
  escapeinside={(*@}{@*)},
  inputencoding=utf8,
  extendedchars=true,
}
}

\definecolor{english}{rgb}{0.0, 0.5, 0.0}

\begin{document}

\title{Accelerated Preference Elicitation with LLM-Based Proxies}

\titlerunning{Accelerated Preference Elicitation with LLM Proxies}

\author{David Huang\inst{1} \and
Edwin Lock\inst{2} \and
Francisco Marmolejo-Cossío\inst{3,4} \and
David C. Parkes\inst{1}}

\authorrunning{Huang et al.}

\institute{Harvard University, USA
\email{davidhuang@seas.harvard.edu, parkes@eecs.harvard.edu}
\and
King's College London, UK
\email{edwin.lock@kcl.ac.uk}
\and
Boston College, USA
\email{marmolf@bc.edu}
\and
Input Output, USA}

\maketitle

\begin{abstract}
Bidders in combinatorial auctions face significant challenges when describing their preferences to an auctioneer. Classical work on preference elicitation focuses on query-based techniques inspired from proper learning---often via {proxies} interfacing between bidders and an auction mechanism---to incrementally learn bidder preferences as needed to compute efficient allocations. Although such elicitation mechanisms enjoy theoretical query efficiency, the amount of communication required may still be too cognitively taxing in practice. 

We propose a family of efficient LLM-based proxy designs for eliciting preferences from bidders using natural language. Our proposed mechanism combines LLM pipelines and DNF-proper-learning techniques to quickly approximate preferences when communication is limited. To validate our approach, we create a testing sandbox for elicitation mechanisms that communicate in natural language. In our experiments, our most promising LLM proxy design reaches approximately efficient outcomes with five times fewer queries than classical proper learning based elicitation mechanisms.

\end{abstract}

\textbf{Code} --- \url{https://github.com/davidzqhuang/alpha}

\section{Introduction}

Auctions are fundamental tools in both theoretical and applied economics, enabling the efficient allocation of resources in domains ranging from spectrum sales \cite{cramton1997fcc} and procurement \cite{klemperer_auctions_2004} to public goods and school assignment \cite{palacios-huerta_combinatorial_2024}. In combinatorial auctions, where bidders express preferences over bundles of items, the burden of fully specifying valuations can be prohibitive. Direct revelation mechanisms require bidders to articulate complete preference reports, which quickly becomes cognitively intractable as the number of items grows \cite{budish2022can,nisan_communication_2006,braziunas_computational_2006}.

Preference elicitation mechanisms aim to mitigate this challenge by incrementally querying bidders, learning just enough about their valuations to determine an efficient allocation. Classical elicitation approaches rely on structured queries—such as value and demand queries—and apply tools from proper learning to incrementally infer bidder types \cite{lahaie2004applying}. While such mechanisms may have polynomial query complexity in theory, they often remain impractical in settings where each query may impose a prohibitive cognitive burden on human participants \cite{ergin2010unique,nielsen_when_2022}.

In response to these challenges, we introduce a novel approach of communicating preferences using natural language. We integrate LLMs into the framework of incremental revelation mechanisms, specifically the competitive equilibrium combinatorial auction (CECA) \cite{lahaie2004applying}. In our approach, LLM-powered proxies act as intermediaries in the elicitation process, interacting with the auction mechanism on behalf of the bidder. The proxies query the bidder through natural language interactions to uncover preferences. The use of LLMs allows the proxy to infer bidder preferences more efficiently than traditional methods.

To evaluate the effectiveness of our proxy designs, we develop a simulation pipeline that models the bidding process in combinatorial auctions. This pipeline makes a second use of LLMs: beyond powering proxies, we use LLMs to simulate human bidders who respond to elicitation queries (cf.~\cite{horton2023large}). Each simulated bidder is defined by a natural language \emph{seed description} that encodes their preferences over bundles. These seeds are then used by a distinct \emph{simulation LLM}, which generates responses to queries—including value queries, demand queries, and open-ended questions—posed by the proxy. 

To validate this framework, we demonstrate that different LLMs produce consistent bundle valuations when conditioned on the same seed, suggesting that our preference modeling is robust across models. This setup enables controlled, repeatable experiments while preserving the complexity and ambiguity inherent in human preference articulation. Throughout our experiments, proxies interact exclusively with these simulation LLMs to elicit and approximate the underlying (implicit) preferences of the bidders they represent.

After the auction concludes, we evaluate the allocative efficiency of the elicited outcome by computing the total social welfare—defined as the sum of the values each simulated bidder assigns to their allocated bundle. We compare this to the welfare-maximizing allocation that could be achieved with full knowledge of the bidders' latent valuations (as implied by the seed-based simulation model). This allows us to assess both the accuracy of the elicitation process and the efficiency of the resulting allocation.

We benchmark our LLM-based proxies against a canonical baseline: the DNF learning proxy from \cite{lahaie2004applying}, which elicits preferences using proper learning from structured queries. Across a range of query budgets, we compare how efficiently each proxy learns bidder valuations and how close the resulting allocation is to optimal. Our results show that the best-performing LLM proxy achieves approximately efficient outcomes using up to five times fewer queries than the classical baseline. These findings demonstrate that natural language–based inference can significantly reduce communication costs in combinatorial auctions while maintaining high-quality allocations.

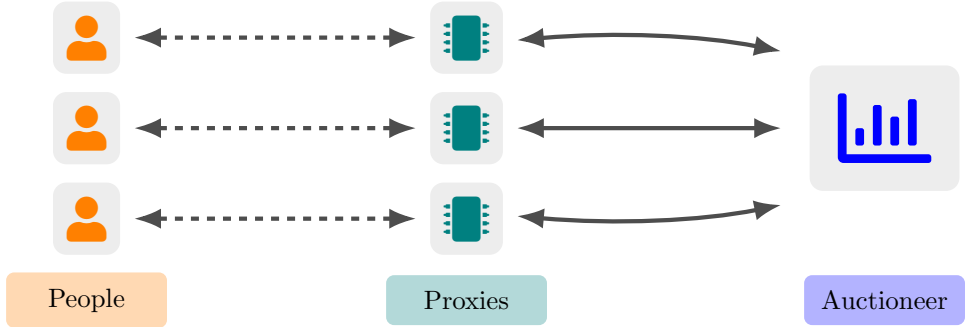
\begin{figure*}[tb!]
\centering
\input{figures/auction-structure.tikz}
\caption{Overview of our proxy auction design with people, proxies and the auctioneer. The auctioneer communicates with the proxies to run the auction, and cannot communicate with the people directly. Each person's proxy maintains a belief of its person's preferences, which it refines over time by communicating with its person using value queries, demand queries and natural language questions.}
\label{fig:auction-setup}
\end{figure*}

\subsection{Related Literature}

Recent work has significantly advanced preference elicitation and efficiency in combinatorial auctions by leveraging machine learning techniques. Approaches such as \textit{probably approximately efficient auctions} \cite{brero2017probably} and \textit{ML-powered iterative combinatorial auctions} \cite{brero2019machine} use bidder-specific models to generalize valuations and improve allocation outcomes. Neural architectures—including \textit{monotone-value neural networks (MVNNs)} \cite{weissteiner2021monotone} and other deep learning frameworks \cite{weissteiner2020deep,soumalias2024machine}—further enhance scalability in high-dimensional settings. Meanwhile, Bayesian methods \cite{brero2019fast,weissteiner2023bayesian} and multi-task learning techniques \cite{maruo2024efficient,soumalias2024machine} support uncertainty modeling and parameter sharing across bidders.

Despite these advances, existing methods operate exclusively over structured input-output formats and lack support for natural language interaction. Our work introduces a complementary approach by using large language models (LLMs) as proxies that elicit bidder preferences through free-form natural language. This requires natural language descriptions of auction scenarios—capabilities not supported by existing benchmarks such as SATS \cite{weiss2017sats}.

Unlike prior work that tests LLMs in narrow or domain-specific elicitation tasks \cite{austin2024bayesian,ronanki2023investigating}, we integrate LLMs with proper learning techniques to enable preference inference through conversational interactions. This combination allows proxies to elicit valuations more intuitively, reducing the cognitive burden on bidders and lowering overall communication overhead. As such, our work presents a novel direction in the design of elicitation mechanisms, bridging classical auction theory with emerging natural language technologies.

Our methodology is closest in spirit to classical preference elicitation via proper learning \cite{lahaie2004applying,blum2004preference}, where queries incrementally recover bidder valuations. We extend this line of work by embedding it within a proxy framework that incorporates LLM-based inference and question generation.
At the same time, our approach continues a long and foundational line of research on competitive equilibrium and auctions, which has traditionally emphasized existence, efficiency, convergence, and computational tractability rather than explicitly modeling misreporting 
(e.g.,~\cite{arrow2024existence,milgrom1982theory,gul1999walrasian,ausubel2004efficient,klemperer2010product,vazirani2007combinatorial}). 
In practice, policymakers often employ auction formats that are not strategy-proof, prioritizing simplicity, transparency, and implementability. Starting from the truthful-reporting setting is therefore both methodologically consistent with this tradition and practically relevant. Our framework extends these insights to settings where preferences are communicated through natural language, introducing a bridge between classical equilibrium theory and modern machine-learning–based elicitation.

This bridge between theory and natural-language elicitation connects our work to emerging efforts that similarly investigate how LLMs can serve as interfaces for preference reporting.
Closely related is concurrent work by Soumalias et al.~\cite{soumalias2025llm}, who study LLM-based elicitation in combinatorial assignment settings. They introduce a system relying on natural language to elicit ranked preferences for simulated users, grounded in real-world domains. As they note, their focus on combinatorial assignment contrasts with our study of combinatorial auctions, which introduces distinct strategic and computational challenges. Their framework does not embed proxies within a formal auction mechanism or benchmark against learning-based baselines. By contrast, our work develops LLM proxies that operate within CECA, evaluates efficiency via welfare comparisons, and introduces a simulation pipeline to analyze elicitation accuracy end-to-end.

\section{The Proxy Auction Model}

We consider a \textit{proxy auction design} with an auctioneer selling goods $G \coloneqq \{g_1, \ldots, g_n\}$, and a population of people $P \coloneqq \{p_1, \ldots, p_m \}$ who each wish to acquire a subset, or \textit{bundle}, of ${G}$. The auctioneer and people do not communicate with each other directly. Instead, each person $p \in P$ is associated with a (distinct) proxy $\omega(p)$, who maintains a belief of the person's preferences and interacts with the auctioneer on her behalf. Figure~\ref{fig:auction-setup} illustrates this setup.

Let $\mathcal{B} \coloneqq \{b \mid b \subseteq G\}$ denote the collection of all bundles. A \textit{XOR bid} $\theta = (B, v)$ consists of a set $B$ of \textit{atomic bundles} satisfying $\{\emptyset\} \subseteq B \subseteq \mathcal{B}$ and \textit{atomic valuation} $v: B \rightarrow \mathbb{R}_+$. Bid $\theta$ induces the valuation $v^{\theta}$ that is defined, for every bundle $b \in \mathcal{B}$, by
\begin{equation}
v^{\theta}(b) \coloneqq \max_{{\{b' \in B \mid b' \subseteq b\}}} v(b').
\end{equation}
So the value of a bundle is determined by the highest-valued atomic bundle it contains. XOR bids can represent any valuation $v$ satisfying the \textit{free disposal} property that $v(b) \leq v(b')$ if $b \subseteq b'$. 

The auctioneer sets prices defined by a \emph{price function} $\varphi^i: \mathcal{B} \to \mathbb{R}_+$ for each person $p_i \in P$. Prices can be non-linear and non-anonymous, so the price assigned a bundle is not necessarily the sum of the prices of its items, and different buyers may face different prices for the same bundle.

The \textit{demand correspondence} $D^{\theta}$ of XOR bid $\theta$ maps every price function $\varphi$ to the set of bundles that maximize quasi-linear utility at these prices, so
\[
D^{\theta}(\varphi) \coloneqq \argmax_{b \in \mathcal{B}} v^{\theta}(b) - \varphi(b).
\]

A \textit{competitive equilibrium} for XOR bids $\theta^1, \ldots, \theta^m$ consists of a price function $\varphi^i$ and allocation $b^i$ for each bid $\theta^i$ such that $b^i \in D^{\theta^i}(\varphi^i)$ and $(b^1, \ldots, b^n)$ partitions the set of goods $G$. Bikhchandani and Ostroy~\cite{bikhchandani2002package} show that a competitive equilibrium is guaranteed to exist for any collection of XOR bids, and maximizes \textit{social welfare} $\sum_{i \in [m]} v^{\theta^i}(b^i)$ among all partitions $(b^1, \ldots, b^m)$ of the goods. The goal of the auctioneer is to find a welfare-maximizing outcome, which he does by computing a competitive equilibrium.

\paragraph{Communication between proxy and person.}
Each proxy maintains a candidate XOR bid, which represents its current belief of its person's preferences. Each proxy gathers information about its person's preferences by making value queries, demand queries, or natural language queries to the person. We write $\mathcal{Q}^p_{N}(q)$ for the answer from person $p$ to natural language query $q$. Similarly, $\mathcal{Q}^p_{V}$ maps any bundle to a (non-negative real) value that person $p$ has for the bundle. Finally, a demand query $\mathcal{Q}^p_{D}$ takes prices $\varphi$ and a bundle $b$ as input and returns either $(1, b)$ if the person is happy with $b$ at $\varphi$ or $(0, b')$ if the person would prefer some bundle $b' \neq b$ to $b$ at $\varphi$. Answers to queries need not be deterministic.  Sections.~\ref{section:DNF-proxy} and \ref{section:LLM-proxies} outline different variations on proxy designs.

Each proxy, $\omega(p)$, maintains a \textit{context} that it updates by repeatedly communicating with $p$. The proxy uses this context to formulate its beliefs of $p$'s preferences. In our LLM-powered proxies, this context includes the \textit{communication history} of queries and responses stored in natural language.\footnote{In our software implementation, the LLM proxies maintain a history containing a transcript of all communications in natural language. Our benchmark DNF learning proxy simply maintains a candidate XOR bid as its context.}

\paragraph{Natural language descriptions.} An integral part of our setup is that a given auction instance, $\mathcal{A}$, is associated with a \textit{description function} that maps $\mathcal{A}$ to a natural language description of the auction itself, its people, and its goods. For example, auction $\mathcal{A}$ might be described by $\mathcal{D}(\mathcal{A}) = $ ``A farmer's market. There are three goods: an apple, a banana, and a strawberry. We have one person, Carlos, who only likes red fruit and has a budget of \$7.''.

The description function of an  auction will be used by our LLM proxy designs, as well as in our LLM pipeline to simulate people in our experiments.

\subsection{The Competitive Equilibrium Combinatorial Auction}
\label{section:CECA}

The auctioneer implements the \textit{competitive equilibrium combinatorial auction (CECA)}~\cite{lahaie2004applying} in our proxy auction setting. In this auction, the auctioneer first asks each proxy for a XOR bid, and then computes a competitive equilibrium for these bids using integer linear programming. The auction then asks the proxies of people not satisfied with their allocation at their prices to update their XOR bids, and repeats the procedure. The proxies thus incrementally elicit new information until a competitive equilibrium satisfying all people can be reached.

Given a XOR bid $\theta$, we define prices $\varphi^{\theta}$ as $\varphi^\theta(b) \coloneqq v^{\theta}(b)$ for each $b \in \mathcal{B}$. These prices support any competitive equilibrium allocation \cite{lahaie2004applying}.\footnote{Although competitive equilibrium allocations are supported by a lattice of prices, we follow \cite{lahaie2004applying} in describing the CECA with maximal competitive equilibrium prices.} 
A formal description of this iterative procedure is provided in Appendix~\ref{appendix:competitive-equilibrium-auction-loop}, but the following is an overview of main steps of the CECA:
\begin{description}[noitemsep]
    \item[Step 1.] Retrieve a XOR bid $\theta^i$ from each proxy $\omega(p_i)$, and compute a competitive equilibrium $(\varphi^{\theta^i}, b^i)$ with maximal competitive equilibrium prices for these bids using integer linear programming.
    \item[Step 2.] Ask every proxy whether their person is satisfied with the allocation $b^i$ at prices $\varphi^{\theta^i}$. If so, \textbf{return} the~competitive~equilibrium.
    \item[Step 3.] Request the proxy of every unsatisfied person to update their XOR bid, and go to Step 1. 
\end{description}

\subsection{Proxy Implementations}
\label{section:message-subroutine}

The CECA is straightforward to implement if all proxies, $\omega$, implement a {\it message subroutine}, $\mathcal{M}^\omega$, that the auctioneer uses to communicate with the proxies. The subroutine takes as input a bundle $b$ and prices $\varphi$. If the person $p$ represented by $\omega = \omega(p)$ is satisfied with bundle $b$ at prices $\varphi$, the subroutine returns $(1, \theta)$, where $\theta$ is the proxy's current candidate XOR bid. Otherwise, the subroutine returns $(0, \theta')$, where $\theta'$ is an updated candidate XOR bid incorporating the new information that the person is unsatisfied with $b$ at $\varphi$. When participants satisfy quasilinear preferences, the proxy implementation of the CECA auction is guaranteed to terminate with the efficient allocation {as long as new information is provided in the updated bid $\theta'$}~\cite{lahaie2004applying} (see also Section~\ref{section:DNF-proxy}). 

\section{The DNF Learning Proxy}
\label{section:DNF-proxy}

We first describe the \emph{\xorproxy{}}, denoted $\omegaXOR$, that is adapted from the preference elicitation literature \cite{lahaie2004applying}. Fix a person $p$. The proxy $\omegaXOR$ maintains a candidate XOR bid $\theta = (B, v)$, initialized as $B = \{\emptyset\}$ and $v(\emptyset) = 0$. It implements the message subroutine $\mathcal{M}^{\omegaXOR}$ from Section~\ref{section:message-subroutine} as the following procedure. 

Given a bundle $b$ and prices $\varphi$ as input, it makes a demand query $Q^p_D(\varphi, b)$ to its person, $p$, and considers the query's outcome $(r, b')$. If $r = 1$, then the person is satisfied with $b$ at $\varphi$, so the algorithm returns $(1, \theta)$. Otherwise, if $r=0$, the algorithm finds a new atomic bundle $b''$ and value $v(b'')$ to add to the XOR bid $\theta$ that it maintains. It then returns $(0, \theta')$, with $\theta'$ being the updated candidate XOR bid.

When searching for a new atomic bundle, $b''$, the proxy queries and stores its person's value, $x$, of the initial bundle $b'$. It then iterates over all items, $i$, in the bundle (e.g., in ascending order) in turn, removing the item from the current bundle $b'$ whenever the value of $b'$ is equal to the value of the bundle $b'$ without item~$i$. Following this loop, the resulting bundle is then returned as $b''$, together with its value, $x$.

This procedure is derived from existing proper learning algorithms for monotone DNF formulas \cite{blum2004preference}. It is immediate that the new atomic bundle $b'$ and its value can be found with at most $n$ valuation queries. With quasilinear preferences, the proxy will  correctly learn the person's valuation in the form of $\theta = (B, v)$ after at most $|B|$ demand queries and $|B|n$ valuation queries. 

The full pseudocode for this procedure can be found in Appendix~\ref{appendix:preference-elicitation-routines}, and we also refer the reader to \cite{lahaie2004applying} for more details. In our experiments, the \xorproxy{} serves as the benchmark from the preference elicitation literature against which we compare our LLM-based proxies.

\section{Our LLM Proxy Designs}
\label{section:LLM-proxies}

We now propose a family of proxy designs that incrementally build upon the DNF learning proxy by incorporating LLMs. As mentioned previously, all our LLM proxies maintain a history of past interactions with their person and the auctioneer in the form of a natural language transcript. This transcript contains all queries made to its person and as well as her answers, and all message requests from the auctioneer and the XOR bids returned by the proxy in response. Additionally, the proxies maintain a belief of their person's preference in the form of a XOR bid $\theta = (B, v)$, which we call the `candidate XOR bid'. The proxy uses their transcript and candidate XOR bid to implement the subroutine $\mathcal{M}^{\omega}$ from Section~\ref{section:CECA}. Throughout this section, we fix a person $p \in P$ for which we describe the proxy, shortening $\omega(p)$ to $\omega$ for convenience.

\subsection{Drop-in LLM Proxies}

\emph{Drop-in LLM proxies} are `drop-in' replacements for proper-learning proxies like $\omegaXOR$ because they communicate with their person using the same message types: value queries and demand queries. We propose two such designs which incrementally build upon the baseline DNF learning proxy.

\paragraph{Design 1.} The LLM proxy $\omegaVDone$ implements the message subroutine $\mathcal{M}^{\omegaVDone}$ as the following procedure:

First it prompts an LLM with the input prices $\varphi$ and allocation~$b$, as well as the auction description $D(\mathcal{A})$ and the proxy's transcript. Then it ask the LLM to choose one of the following actions:
\begin{enumerate}[noitemsep]
\item Indicate that the person is satisfied;
\item Message the person a value query;
\item Message the person a demand query.
\end{enumerate}

If the LLM chooses action 1, it returns $(1,\theta)$, where $\theta$ is proxy's current candidate XOR bid.
For action 2, the LLM is also asked to provide a bundle $b$ with which to augment the proxy's candidate XOR bid. The proxy then makes the query $\mathcal{Q}^p_V(b)$ to its person $p$ to ascertain her value, $z$, for bundle $b$, and extends $\theta = (B,v)$ to a new XOR bid $\theta'$ by adding $b$ to $B$ and letting $v(b) = z$. Subsequently, the proxy returns $(1, \theta')$.
For action 3, the proxy make a demand query $\mathcal{Q}_D^p(\varphi, b)$ and retrieves a bundle $b'$ demanded by $p$. If $b' = b$, then it returns $(0, \theta)$. Otherwise, it makes the query $\mathcal{Q}_V^p (b')$ to ascertain person $p$'s value, $x$ for $b'$ and returns $(1, \theta')$, where $\theta'$ is the augmented XOR bid as above.

\paragraph{Design 2.} The second drop-in proxy $\omegaVDtwo$ is similar to $\omegaVDone$. As with $\omegaVDone$, it maintains an internal candidate XOR bid, $\theta' = (B', v')$ as a part of its context. Unlike $\omegaVDone$, the proxy $\omegaVDtwo$ reports to the auctioneer an {\it inferred} XOR bid $\theta^* = (B^*, v^*)$, which agrees with $\theta'$ for all bundles $b \in B'$, and additionally infers values $v^*(b)$ for all $b \in \mathcal{B} \setminus B'$.

$\omegaVDtwo$ computes $\theta^*$ using an inference function $\gamma: \mathcal{B} \rightarrow \mathbb{R}_+$ that gives an inferred value for each bundle. To do this, the proxy follows the following procedure: it maintains a working $\theta^* = (B^*, v^*)$ initialized to $\theta'$, and iterates over all possible bundles in ascending order of the number of items they contain and for each bundle $b$, if $b \notin B'$ and $v^*(b) < \gamma(b)$, it adds $b$ to $B^*$ and defines $v^*(b) = \gamma(b)$.

The inference function $\gamma$ is calculated at scheduled iterations of the CECA auction, including the first iteration, using an LLM pipeline. The LLM, given the proxy's transcript, the auction description $D(\mathcal{A})$, and a bundle $b$, infers a value $z$ for the bundle. Thus $\gamma$ can be constructed by calling the pipeline with all $b \in \mathcal{B} \setminus B'$. 
Finally, $\omegaVDtwo$ discounts the value given by the LLM by a constant factor of $\epsilon = 0.75$ to prevent $\omegaVDtwo$ from over-reporting its person's valuation for bundles. The reason for this is two-fold: 1) over-estimation of valuation can lead to artificially high prices for bundles which a person could be allocated otherwise, and 2) artificially high prices can result in a person over-paying and obtaining negative utility from participating in the auction. Moreover, $\varepsilon = 0.75$ is set so that the proxy does not over-bid the person's value. 

\subsection{Plus-Questions LLM Proxy}

The \textit{plus-questions proxy} $\omegaNVD$ uses natural language questions in addition to value queries and demand queries when messaging their person. The proxy implements $\mathcal{M}^{\omegaNVD}$ almost identically to the implementation of $\omegaVDtwo$, except that the very first time it is run, it first calls $\mathcal{Q}_N^{p}(q)$, where question $q$ is given by the LLM response to a prompt asking, ``What should the proxy ask the person next to better understand their preferences?" given the auction description $\mathcal{D}(\mathcal{A})$.

Note that $\omegaNVD$, like $\omegaVDtwo$, also returns an inferred XOR bid $\theta^*$. The additional information that person $p$ provides in response to $q$ is recorded in the proxy's transcript, and thus is used in all subsequent calls to the message subroutine to give more information to the LLM when it chooses which action to take, and how to infer its new candidate XOR bid.

\subsection{Hybrid Proxy}

The \emph{hybrid proxy} $\omegaH$ combines $\omegaNVD$ and the $\omegaXOR$. The procedure implementing $\mathcal{M}^{\omegaH}$ has two parameters, $\alpha \in \mathbb{Z}_+$ and $\delta \in (0,1)$. For the first $\alpha$ calls to $\mathcal{M}^{\omegaH}$, the procedure simulates the procedure of $\omegaNVD$ that implements $\mathcal{M}^{\omegaNVD}$. So the proxy initially asks a natural language question and uses its LLM to infer valuations. After $\alpha$ iterations, the procedure instead simulates the procedure of $\omegaXOR$ implementing $\mathcal{M}^{\omegaXOR}$ and thus uses the proper-learning approach from the preference elicitation literature. However, recall how $\mathcal{M}^{\omegaVDtwo}$ periodically recomputes the inference function $\gamma$ and, on each call, calculates and return the inferred valuation $\theta^*$. The procedure of $\omegaH$ continues to derive an inferred valuation after simulating the procedure of $\omegaXOR$. We specify that in every iteration after the switchover, we update $\gamma$ to $\gamma' = \delta \gamma$. Multiplying with $\delta$, which lies in $(0,1)$, has the purpose of maintaining, in the short run, the benefits of using an inferred valuation, but in the long run, removing the inferred valuation by decaying $\gamma$ towards~$0$.%
\footnote{In our experiments, the parameters $\alpha$ and $\delta$ are selected through inspection and empirical validation during proxy development; $\alpha$ is set to $10$ because the diminishing return to efficiency of the LLM proxies is readily apparent (Figure~2); $\delta$ is set to $0.95$ for it to be sufficiently high enough to facilitate a smooth transition from the LLM to the XOR proxy's bid, but low enough to decay the inference function mappings to $0$ within a number of steps similar to that the XOR proxy takes to reach 100\% efficiency.}

\section{The LLM Simulation Pipeline}

In order to evaluate the effectiveness of the \xorproxy{} and our LLM proxies, we {make another use of LLMs, also designing} an LLM pipeline for simulating people in our auction setup. This is used in Section~\ref{section:experiments} for our experiments. In our simulations, the LLM answer natural language questions as well as value and demand queries. The approach builds on recent work demonstrating the utility of LLMs for modeling human behavior in economic settings~\cite{brand2023using,horton2023large}. While the realism of our results ultimately depends on the fidelity of the LLM as a model of human responses, we validate the robustness of the simulation in several ways, as detailed in Section~\ref{subsec:robustness}.

Preferences are developed by our simulation pipeline by constructing a \textit{seed} description, $\mathcal{D}(p)$, in natural language for each simulated person $p$. These seeds are crucial for the simulated person to be able to answer natural language questions about their preferences.
This contrasts with other methods of generating random preferences, e.g., simulated preferences that arise from arbitrary orderings of bundles. Such methods do not necessarily represent realistic preferences and, secondly, cannot be described meaningfully in natural language.

\subsection{Seed Generation LLM Pipeline}
\label{subsec:seed-generation-pipeline}

The process for generating the seeds, $\mathcal{D}(p)$, that define person $p$'s preferences consists of a four-step LLM pipeline. The first step generates the germ of the seed, and the three subsequent steps revises the seed to be more specific and more coherent. We provide example seeds generated by this procedure in Appendix~\ref{appendix:example-seed}. 

\begin{description}
    \item[Step 1: Initial preference generation.] The LLM generates a person's specific purchase occasion and taste over the items, providing an initial description of a person's preferences based on the given auction scenario. To encourage diversity in responses, we ask the LLM to generate three possible versions and then randomly sample one.
    \item[Step 2: Revision for clarity and consistency.] The LLM revises the initial preferences to enhance clarity and internal consistency in describing the person's preferences.
    \item[Step 3: Anchor with specific values.] The LLM anchors the person's preferences by assigning explicit dollar values to specific bundles of interest.
    \item[Step 4: Refinement for precision in complex bundles.] The LLM refines the person's preferences to outline a process that defines how they evaluate other unspecified bundles, considering items as substitutes and complements.
\end{description}

\subsection{Message Response LLM-Based Pipeline}

In order to ensure that the simulated person acts coherently, consistent with a standard model of preferences, the pipeline initially constructs an XOR bid $\theta = (B, v)$ for the person, which is then cached. The pipeline then responds to value and demand queries using $\theta$.

We implement a subroutine to perform LLM calls and help construct $\theta$. This subroutine takes as input a bundle and returns an \textit{initial} value for this bundle. Maintaining a working valuation $\theta = (B, v)$, we iterate over all possible bundles in ascending order of the number of items they contain. At each iteration, we run the first subroutine to get the \textit{initial} value $\nu$ of the bundle and update $\theta$ with that bundle and its value if free disposal is not violated, i.e, if $v^*(b) < \nu$. 
Details regarding the prompt pipeline for value queries are in Appendix \ref{appendix:bundle-valuation-subroutine}. Prompt pipelines for natural language and equivalence queries are similar in nature, and can be found in Appendix \ref{appendix:natural-language-questions} and Appendix \ref{appendix:equivalence-query-subroutine} respectively.


\begin{table*}[tb!]
\setlength{\tabcolsep}{3pt} 
\centering
\footnotesize
\begin{tabular}{p{.31\linewidth}p{.34\linewidth}p{.32\linewidth}}
\toprule
\textbf{Electronics} & \textbf{Preserves} & \textbf{Transportation} \\
\midrule
A collection of like-new electronics donated to a local {library}, ranging from AirPods to iPads and accessories.
&
Gourmet food items are available, including exotic jams and chutneys with unique flavors, perfect for cooking or gourmet meals.
&
A variety of electric scooters and bikes are up for auction, catering to urban commuters and fitness enthusiasts.
\\
\cmidrule{1-3}
Apple AirPods (2nd Gen) & Organic Strawberry Jam & Electric Scooter S2 Pro (2024) \\
Apple AirPods Max & Wild Blueberry Preserves & \mbox{E-Scooter~Elettrica~(2023)} \\
Apple iPad (9th Gen) & Apricot \& Lavender Conserve & \mbox{Voltron SP03 E-Scooter (2024)} \\
Apple iPad Air (M2) & Sugar-Free Raspberry Spread & Troik Verve+ 2 (2023) \\
Apple Pencil (2nd Gen) & Spiced Plum Chutney & Titan Escape 3 (2023) \\
Apple Pencil Pro & Mango \& Passionfruit Jam & Schwin Suburban (2021) \\
 \bottomrule
\end{tabular}
\caption{The three auction descriptions and items used in our experiments.}
\label{table:scenario-items}
\end{table*}

\section{Experiments}
\label{section:experiments}

The experiments, along with the simulation toolbox, were implemented in Python 3. We used Google's OR-Tools package for integer linear programming for the auction. OpenAI's gpt-4o-mini model (versioned to July 18, 2024) is used as our LLM for the proxy and the simulation pipeline.

The three auction scenarios we use in our experimental evaluation of our LLM proxies are outlined in Table~\ref{table:scenario-items}. They are designed be accessible so that the reader can readily assess the realism of the interactions and various LLMs can be tested without domain-specific knowledge or fine tuning. Each scenario has $N = 6$ items for auction. Some items are clear substitutes and others are clear complements.

For each scenarios, we generated three tranches of three seeds each, for a total of $27$ simulated participants in the auctions.

In Sections~\ref{subsec:robustness} and \ref{subsec:preference-profile}, we run experiments on the LLM simulation pipeline to validate the robustness and sensibility of the preferences of the simulated people created by the pipeline. In Sections~\ref{subsec:auction-efficiency} and \ref{subsec:hybrid}, we run experiments to test the performance of our LLM proxy designs against the benchmark \xorproxy{} design.

\subsection{Auction Efficiency}
\label{subsec:auction-efficiency}

\begin{figure}[tb!]
\centering
\includegraphics[scale=0.5]{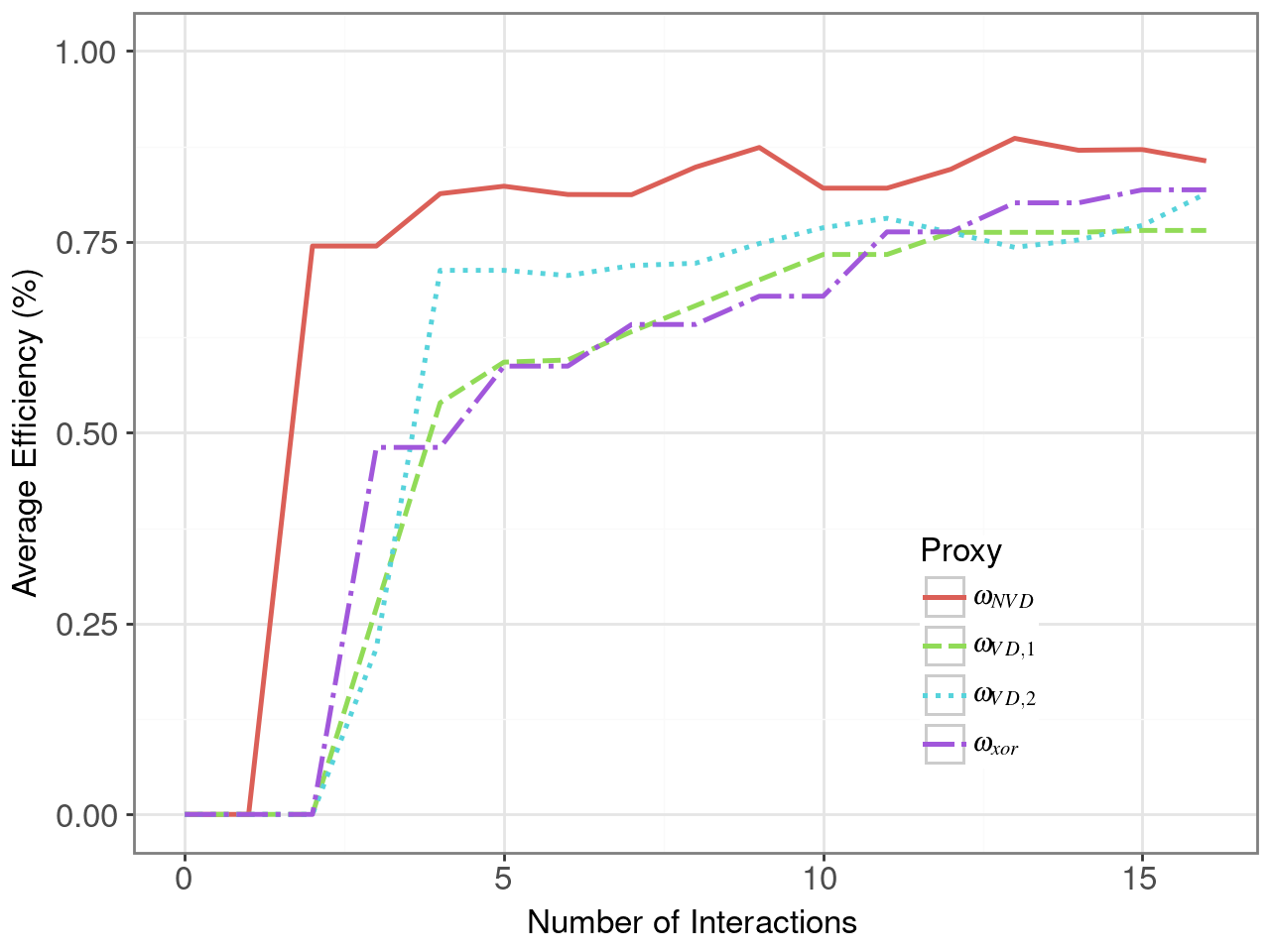}
\caption{Average efficiency of the auction run under the three scenarios from Table~\ref{table:scenario-items}. Anchoring on our \xorproxy{}, we see that the drop-in LLM proxy $\omegaVDone$ performs similarly, also reaching 75\% efficiency in around $10$ person--proxy interactions. The proxy $\omegaVDtwo$ performs significantly better, reaching 75\% efficiency in four interactions;  proxy $\omegaNVD$, which additionally uses a natural language question, reaches 75\% efficiency in two interactions.}
\label{fig:auction-value}
\end{figure}

We explore a set of simulations of auctions to see how quickly LLM proxies (drop-in and plus-questions proxies) reach efficient outcomes, compared to the \xorproxy{}. Our simulations show that efficient outcomes can be reached with less communication using LLM-based proxies. Quantitatively, we can reach 75\% efficiency in ten interactions person-proxy interactions with $\omegaXOR$, in four interactions with $\omegaVDtwo$, and in two interactions with $\omegaNVD$ (see Figure~\ref{fig:auction-value}).

For the nine tranches of three people each with three tranches per scenario, we run the auction using either $\omegaXOR$, $\omegaVDone$, $\omegaVDtwo$, or $\omegaNVD$ for all people, halting the simulation after $16$ iterations. We fully elicit each of the $27$ people's preferences across all $2^6$ bundles, and use that to calculate the welfare-maximizing outcome.

Each iteration of the competitive equilibrium auction, we calculate the social welfare of the auction and record the number of interactions between each person and their proxy. We then calculate, for each iteration, the percent efficiency by dividing the social welfare by that of the social welfare maximizing outcome. We also calculate the average number of interactions. We then determine for $i = 1, 2, \dots, 16$ the efficiency with $i$ interactions by selecting the efficiency at the iteration of the run that has the lowest number of interactions greater than $i$. Then, for each $i$, we average over all nine tranches to get the percent average efficiency shown in Figure~\ref{fig:auction-value}.

\subsection{Long-Run Performance of Hybrid Proxy}
\label{subsec:hybrid}

Section~\ref{subsec:auction-efficiency} shows how LLM-based proxies outpace the proper-learning-based $\omegaXOR$ in the short run. We now consider the hybrid proxy $\omegaH$ to characterize the long-term performance of an LLM-based proxy, showing that we can combine the short-term rapid increase in efficiency without sacrificing the long-term ability to exactly learn the person's valuation and to converge to the social-welfare-maximizing auction outcome. For the following experiments, we set the hybrid proxy parameters to $\alpha = 10$ and $\delta = 0.95$.

First, we run $\omegaH$ and $\omegaXOR$ as preference learning algorithms, by running the auction with only one person with either $\omegaH$ or $\omegaXOR$ as a proxy, constantly setting prices equal to the candidate XOR bid values and allocating the empty bundle, and measure at each step the number of times the proxy interacts with the person and the average error, calculated as the average absolute difference between the value of an item calculated using the person's XOR bid and the value of an item calculated using the proxy's candidate XOR bid. These numbers are aggregated over the multiple tranches using the methodology in Section~\ref{subsec:auction-efficiency}. 

\begin{figure}[tb!]
\centering
\includegraphics[scale=0.125]{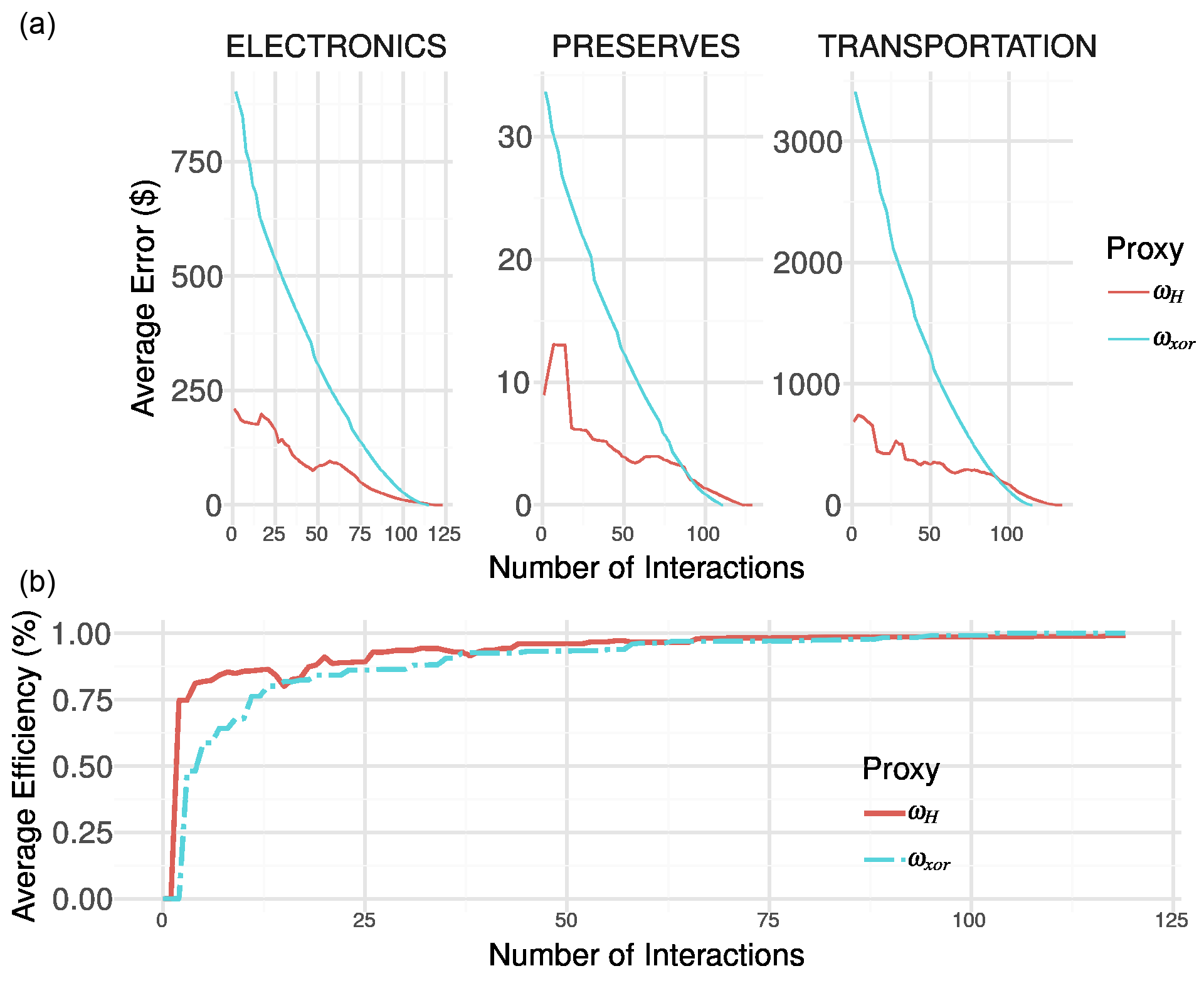}
\caption{Performance of the hybrid proxy $\omegaH$ and the \xorproxy{} $\omegaXOR$ with respect to number of person--proxy interactions. Panel (a) shows that $\omegaH$ achieves a good approximation of its person's valuation significantly quicker than $\omegaXOR$, while still converging to the person's exact valuation. Panel (b) shows a rapid efficiency increase initially for $\omegaH$, and similar long-run efficiency.}
\label{fig:elicitation}
\end{figure}

The results, plotted in Figure~\ref{fig:elicitation} (a), show that $\omegaH$ consistently approximates the person's valuation with fewer interactions. In all three scenarios, $\omegaH$ achieves a better approximation of the person's valuation than $\omegaXOR$ does in $50$ interactions. We note, however, that at the very end, $\omegaXOR$ obtains $0$ error, so it exactly learns the person's valuation with approximately $10$ fewer iterations. This is because the LLM is sometimes inefficient with its queries, and natural language questions are more helpful for approximations than exact learning.

Second, we ran the auction simulations identical to those in Section~\ref{subsec:auction-efficiency}, except that we let the auction continue running until completion. We aggregate and plot the data according to the same procedure and plot the results in Figure~\ref{fig:elicitation} (b). In the initial phase, $\omegaH$ rapidly reaches fairly efficient outcomes. This matches the results shown for $\omegaNVD$. This is expected, as $\mathcal{M}_C^{\omegaH}$ simulates $\mathcal{M}_C^{\omegaNVD}$ in the first $\alpha$ iterations. We also see that the efficiency of the outcome steadily increases as we increase the number of interactions, in line with $\omegaXOR$, so we reach an outcome that maximizes social welfare.

\subsection{Robustness of LLM-Based Person Modeling}
\label{subsec:robustness}

\begin{figure}[tb!]
  \centering
  \includegraphics[scale=0.55]{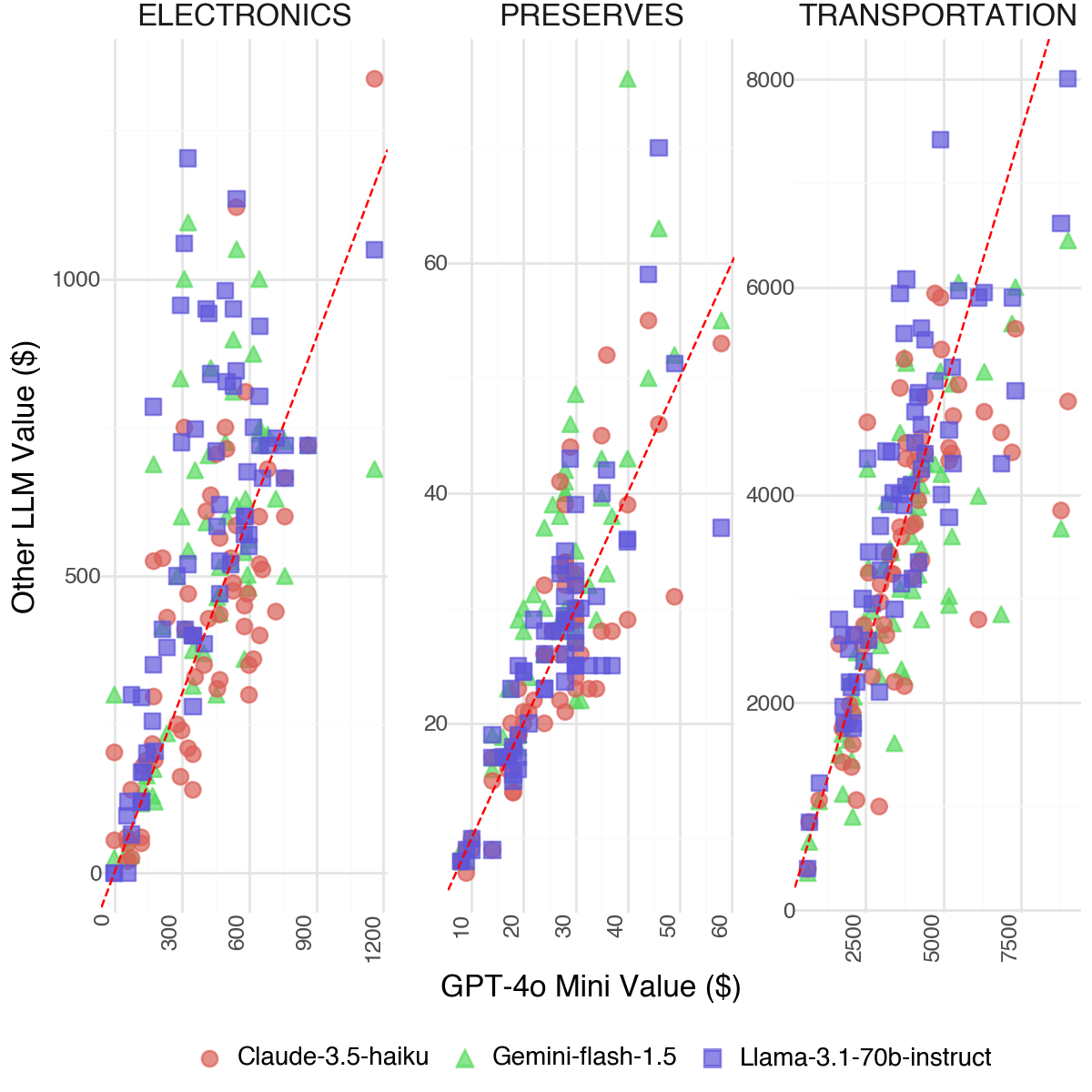}
  \caption{Comparison between GPT-based and other LLM models models in person robustness tests. Gpt-4o-mini and other LLM-models give similar values in response to value queries evidenced by the dots, each of which representing a single bundle, falling near the identity gpt-4o-mini value (\$) = Other LLM value line (\$).}
  \label{fig:gpt-v-gemini}
\end{figure}

We find that the seeds given by the seed-generation LLM pipeline of the simulation are robust. To understand whether the seeds robustly describe a specific valuation, we examine the inter-model robustness of the seeds, comparing the valuations generated by the message response LLM-pipeline using the gpt-4o-mini model to those generated with the gemini-1.5-flash model,
we find that the valuations are broadly consistent, as shown in Figure~\ref{fig:gpt-v-gemini}. As the models are developed independently by OpenAI and Google, respectively, this shows that the seeds, and the valuations interpreted from them, via the message-response pipeline are not specific to a specific family of models and thus robust.

\subsection{Coherent Preferences}
\label{subsec:preference-profile}

We find that the preferences created by the LLM simulation pipeline are coherent. For this, we investigate the preferences generated by the pipeline by running the following experiment. For a specific seed description in each scenario, start with the empty bundle and first incrementally add items until the bundle contains all items, before incrementally removing items again until the bundle is empty. After each addition and removal, we run the message response LLM pipeline to perform a value query, repeating this ten times to account for variability arising from randomness.

\begin{figure}[tb!]
  \centering
  \includegraphics[scale=0.5]{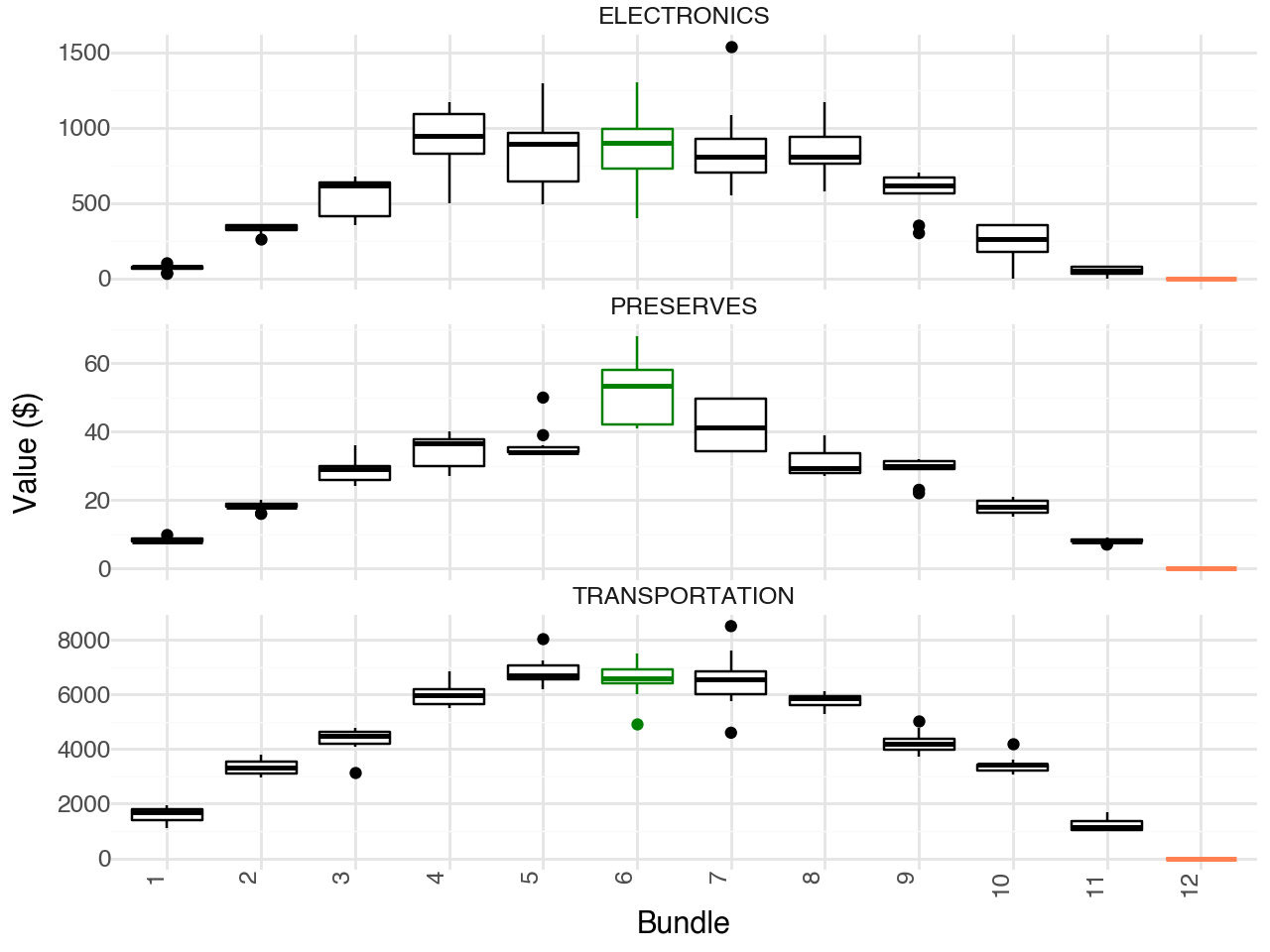}
  \caption{Preference variability and shape after iteratively adding items to an initially empty bundle, and then iteratively removing items. Each box plot displays the distribution of values obtained from with 10 iterations of the process.}
  \label{fig:variability}
\end{figure}

We find that the valuations given are precise. The variation for a given bundle is tight and the overall shape satisfies desirable properties such as monotonicity and free disposal. Figure~\ref{fig:variability} shows that the value of the bundle consistently increases as we add items to the bundle, and decreases as we remove items. 

\section{Conclusion}

This study presents a novel approach to preference elicitation in combinatorial auctions by integrating Large Language Models (LLMs) into proxies. Traditional query-based elicitation methods, while theoretically efficient, often impose substantial cognitive burdens on people.
Our LLM-powered proxy design mitigates this through natural language interactions and inference, leading to more human-centric and efficient auction mechanisms in increasingly complex markets.

Our experiments demonstrate that LLM-based proxies, particularly those that use LLMs to infer the person's valuation on unspecified bundles, achieve approximately efficient auction outcomes up to five times faster than classical DNF learning proxies. As illustrated in Figure~\ref{fig:auction-value}, proxies such as $\omegaVDtwo$ and $\omegaNVD$ reach 75\% of maximum social welfare with markedly fewer interactions compared to the $\omegaXOR$ proxy. Furthermore, the hybrid proxy $\omegaH$ not only accelerates the initial elicitation phase but also converges to the social-welfare-maximizing outcome, suggesting a path forward for combining the rapid approximation capabilities of LLMs with the guarantees of classical preference elicitation.

Additionally, we validate the robustness of LLM-based person modeling across different models and repeated trials, ensuring consistent and reliable preference simulation. This robustness underscores the feasibility of employing natural language processing in complex auction settings, enhancing both the theoretical and practical aspects of auction design.

In summary, our findings highlight the potential of LLMs to improve preference elicitation in combinatorial auctions by making the process more intuitive and less burdensome by using natural language questions and reducing the number of interactions for agents. In addition, we contribute a comprehensive testing sandbox for simulating bidders in auction environments that use natural language both to describe the auction and to communicate between auction participants and proxies.

In future work, the practicality and efficacy of our LLM proxies can be explored in lab experiments featuring human participants representing bidders instead of our LLM-powered simulation pipeline. Further research could also examine the scalability of this approach to larger auction environments and investigate the integration of increasingly advanced LLM capabilities to further enhance elicitation efficiency and accuracy at lower cost. A complementary direction concerns strategic behavior and adversarial communication. While our present analysis assumes truthful communication between bidders and their proxies, extending the framework to account for strategic misreports, adversarial natural-language deviations, or LLM-induced errors would be highly valuable. The unstructured nature of natural language makes it non-trivial even to define profitable deviations, which may in turn dampen incentives for manipulation. Exploring these dynamics could yield new insights into strategy-robust design and deepen our understanding of equilibrium behavior in natural-language elicitation mechanisms.

\bibliographystyle{splncs04}
\bibliography{refs}

\begin{credits}
\subsubsection{\ackname} 
We would like to thank the anonymous reviewers for their thoughtful and constructive feedback, which helped improve the quality and clarity of this work. We also appreciate the valuable discussions and suggestions from colleagues and collaborators that informed various aspects of this research.

\end{credits}

\appendix

\section{Example Seed}
\label{appendix:example-seed}

As part of our simulation methodology, we generate natural language seeds that encode a person's underlying preferences over bundles. These seeds serve as the foundation for how the simulation LLM responds to queries. Below, we present an example seed generated by the four-step LLM pipeline from Section~\ref{subsec:seed-generation-pipeline}:

\begin{MyVerbatim}
Cecilia is focused on acquiring the Apple Pencil Pro (code: APPLEPENCILPRO) for her digital art. She places a high value on tools that enhance her creative capabilities, specifically seeking industry-leading performance and precision for her artistic projects. The dollar value for the Apple Pencil Pro she is willing to pay is estimated at $120, as this aligns with retail market prices for similar tools.

Cecilia is open to considering additional items that can complement the Apple Pencil Pro and significantly improve her workflow. Among the other available items, she may find the following combinations valuable, each associated with an estimated dollar value based on market rates:

1. **Single Item Bundle:**
   - Apple Pencil Pro (code: APPLEPENCILPRO) - $120

2. **Complementary Bundles:**
   - iPad Air 11-inch (M2) with Apple Pencil Pro:
     - iPad Air (code: IPAD12) - $600
     - Apple Pencil Pro (code: APPLEPENCILPRO) - $120
     - Total Value: $720
   - iPad (9th Generation) with Apple Pencil Pro:
     - iPad (code: IPAD9) - $350
     - Apple Pencil Pro (code: APPLEPENCILPRO) - $120
     - Total Value: $470

Cecilia considers these other items because they can enhance her digital art experience, allowing her to utilize the Apple Pencil Pro more effectively. The iPad models provide her the necessary platform for creative applications, which is a key factor in her evaluation.

However, Cecilia prioritizes the Apple Pencil Pro as her primary target and feels that any additional item should not just be an upgrade but must genuinely elevate her creative process. She is likely to discount bundles that include items she believes do not contribute directly to her artistic workflow, such as the Apple AirPods or AirPods Max, as they do not enhance her art-making activities. 

In conclusion, while Cecilia's main interest is the Apple Pencil Pro, she values potential bundles that combine it with compatible iPads, estimating these combinations could be worth between $470 and $720 depending on the iPad model chosen, thus weighing her bids according to the increased functionality they provide for her creative work. 
Cecilia's evaluation process for complex bundles involves a systematic assessment of how each item's inclusion enhances her creative workflow. She prioritizes the Apple Pencil Pro as her core target, and her perception of any additional items is predominantly based on their ability to provide meaningful improvements to her digital art experience.

When considering large bundles, Cecilia does not view them merely as additive collections of items; she assesses the synergy between constituents. Each additional item must deliver substantial value to her artistic capabilities; otherwise, she perceives them as superfluous and potentially detracting from her primary objective. Therefore, while she appreciates bundles that combine compatible products (like the iPads with the Apple Pencil Pro) for their holistic benefits, she remains cautious with larger bundles that include irrelevant components, which she would discount heavily or avoid.

Regarding the necessity of large bundles, Cecilia is not averse to accepting items individually; rather, she prefers a tailored approach that allows her to select only those products that meet her artistic needs. However, if a bundle includes items she considers essential or significantly enhancing (like the iPad models), she is more likely to accept it despite its size.

Cecilia values multiples of similar items additively, but she applies a discount for redundancy. For example, if two artistic tools serve similar functions, she might estimate their combined value at 80% of the total individual retail prices, reflecting a 20% discount on the perceived benefit of repeated functionality. Her evaluation criteria incorporate percentage discounts based on utility:

1. **Complementary Value:** Items that enhance functionality (like the iPads with the Apple Pencil Pro) are valued at a premium, with a direct uplift of about 20% to 30% in her willingness to pay.
2. **Substitution Effects:** If an item is viewed solely as a substitute rather than an enhancement (e.g., she may consider less advanced versions or unnecessary accessories), she would apply a significant discount, potentially around 50% or more on her perceived value.
3. **Bulk Effects:** For larger bundles, Cecilia's assessment would involve a progressive discount model; the first two items may maintain nearly full value, but each subsequent item may see a decline in perceived value of about 10% to 15%, reflecting the diminishing returns of utility in a larger collection.

In conclusion, Cecilia's decision-making process revolves around a clear framework where she rigorously evaluates the functional contributions of each item based on how effectively they can support and amplify her artistic endeavors. This meticulous approach ensures that her bids are strategically aligned with her artistic goals, emphasizing quality and relevant enhancements over quantity.
\end{MyVerbatim}

\section{Algorithmic \& Optimization Details}
\label{appendix:algorithms}

The material collected in this appendix spells out every computational piece that drives the auction-framework described in Section~\ref{section:CECA}. The algorithms are grouped in the order they are used at run-time:

\begin{enumerate}
  \item \textbf{Preference-elicitation routines}  
        (Algorithms~\ref{alg:learn_xor}--\ref{alg:learn_step})  
        learn an \emph{XOR bid} from scratch.  
        Starting from the empty hypothesis, the outer loop repeatedly challenges its current guess with a demand (or equivalence) query; whenever a bidder objects, the \emph{step} sub-routine trims the returned bundle to the smallest \emph{atomic} bundle that still carries the same value.  
        Iteration stops exactly when every further challenge is rejected, guaranteeing that the discovered set of atomic bundles reproduces the bidder's valuation on \emph{all} bundles.\footnote{The procedure is a monotone-DNF learner in disguise.}

  \item \textbf{Competitive-equilibrium auction loop}  
        (Algorithm~\ref{alg:run_ceca})  
        is the high-level CECA driver.  
        At every round it (i) solves the current \emph{winner-determination problem} to obtain a welfare-maximising tentative allocation, (ii) posts bundle-specific \emph{Lindahl prices} that would make that allocation a competitive equilibrium if bids were final, and (iii) lets each proxy either accept the outcome or supply a refined bid.  
        Convergence is provably finite because every rejection enriches at least one manifest valuation.

  \item \textbf{Winner-determination formulations}  
        Equations~\eqref{eq:wdp} and~\eqref{eqs:ilp} cast social-welfare maximisation as a 0-1 integer programme.  
        The binary variable $x_{i,j}$ flags whether bidder~$i$ wins her $j$-th atomic bundle; the first constraint ensures at most one atomic bundle per bidder (respecting XOR semantics), the second guarantees that no physical item is sold twice.  
        Feeding this model to a modern MIP solver (we use OR-TOOLS' SCIP backend) yields the optimal intermediate allocation in milliseconds for the instance sizes considered in our experiments.

  \item \textbf{Lindahl pricing rule}  
        Equation~\eqref{eq:lindahl-prices} sets bidder-specific, bundle-specific prices that support any welfare-maximising allocation as a competitive equilibrium.  
        Intuitively, each bidder pays her own reported marginal contribution above a bidder-specific constant $d_i$; choosing $d_i=0$ produces the \emph{maximal} price vector we adopt in all simulations.
\end{enumerate}

Together, these components provide a complete, constructive path from raw natural-language preferences to an economically efficient, incentive-compatible allocation, and give concrete hooks for the LLM-based proxies presented.

\subsection{Preference-Elicitation Routines}
\label{appendix:preference-elicitation-routines}

\begin{algorithm}[htb!]
\caption{Learn XOR Preferences (\( \mathcal{L}_{xor}(p) \))}
\label{alg:learn_xor}
\begin{algorithmic}[1]
\STATE \textbf{Input:} Person \( p \)
\STATE \textbf{Initialize} \( B \gets \emptyset \), \( v \gets \emptyset \), equivalent \( \gets \) \textbf{False}
\WHILE{not equivalent}
    \STATE \( (r, b) \gets \mathcal{M}^p_{D}(\phi^*_{(B,v)}, b) \) \COMMENT{Obtain response and bundle from demand query}
    \IF{\( r = 1 \)}
        \STATE equivalent \( \gets \) \textbf{True}
    \ENDIF
    \IF{\( r = 0 \)}
        \STATE \( (b', \text{value}) \gets \) \( \mathcal{L}_{xor, step} ( p, b ) \)) \COMMENT{Identify atomic bundle and its value}
        \STATE \( B \gets B \cup \{b'\} \)
        \STATE \( v(b') \gets \text{value} \)
    \ENDIF
\ENDWHILE
\STATE \textbf{return} \( (B, v) \)
\end{algorithmic}
\end{algorithm}

\begin{algorithm}[htb!]
\caption{Learn XOR Preferences Step (\( \mathcal{L}_{xor, step} ( p, b ) \))}
\label{alg:learn_step}
\begin{algorithmic}[1]
\STATE \textbf{Input:} Person \( p \), Bundle \( b \) \COMMENT{Current bundle from demand query}
\STATE \( \nu \gets \mathcal{M}^p_{V}(b) \) \COMMENT{Obtain original bundle value}
\STATE \( b' \gets b \) \COMMENT{Initialize refined bundle}
\FOR{\( i = 1 \) to \( N \)}
    \IF{\(T_i \in b'\)}
        \STATE \( b'' \gets b' \setminus \{T_i\} \) \COMMENT{Remove item \( i \) from bundle}
        \STATE \( \nu' \gets \mathcal{M}^p_{V}(b'') \) \COMMENT{Query new bundle value}
        \IF{\( \nu' = \nu \)}
            \STATE \( b' \gets b'' \) \COMMENT{Item \( i \) is not essential}
        \ENDIF
    \ENDIF
\ENDFOR
\STATE \textbf{return} \( (b', \nu) \) \COMMENT{Return identified atomic bundle and its value}
\end{algorithmic}
\end{algorithm}

\newpage
\subsection{Competitive-Equilibrium Auction Loop}
\label{appendix:competitive-equilibrium-auction-loop}

\begin{algorithm}[!htb]
\caption{Competitive Equilibrium Combinatorial Auction (\( \mathcal{L}_{CECA} \))}
\label{alg:run_ceca}
\begin{algorithmic}[1]
\STATE \textbf{Input:}
\STATE \quad \( \mathcal{S} \): Scenario
\STATE \quad \( p'_1, p'_2, \dots, p'_K \): Proxies for each person $p_i$
\STATE \textbf{Initialize:}
\STATE \quad Set manifest valuations \( \tilde{\theta}_1, \tilde{\theta}_2, \dots, \tilde{\theta}_K \gets (\emptyset, \emptyset \rightarrow 0)\) \COMMENT{Initial guess, e.g., all zeros}
\STATE \quad Set \( \text{counterexample\_flag} \gets \text{False} \)

\WHILE{not \( \text{counterexample\_flag} \)}
    \STATE \quad \( b_1, \dots, b_K \gets \text{allocate}(\mathcal{S}, \tilde{\theta}_1, \tilde{\theta}_2, \dots, \tilde{\theta}_K) \) \COMMENT{Allocate bundles to maximize social welfare via integer programming}
    
    \STATE \quad \( \text{counterexample\_flag} \gets \text{False} \)
    \FOR{each proxy \( i = 1 \) to \( K \)}
        \STATE \quad \( \phi_i(b_i) \gets \text{Compute maximal Lindahl prices}(\phi, b_i) \) \COMMENT{Per Equation~\ref{eq:lindahl-prices}}
        
        \STATE \quad \( (r, \theta') \gets \mathcal{M}^{p_i'}_{C}(\phi_i, b_i) \) \COMMENT{Check if happy}
        
        \IF{\( r = 0 \)}
            \STATE \quad \( \theta_i \gets \theta' \)
            \STATE \quad \( \text{counterexample\_flag} \gets \text{True} \) \COMMENT{Need to reallocate}
        \ENDIF
    \ENDFOR
\ENDWHILE
\STATE \textbf{return} \( \{ (b_i, \phi_i(b_i)) \mid i = 1, 2, \dots, K \} \) \COMMENT{Allocated bundles and corresponding payment}
\end{algorithmic}
\end{algorithm}

\subsection{Winner-Determination Formulations}
\label{appendix:winner-determination-formulations}

\paragraph{Social-Welfare Objective}
\begin{equation}
\label{eq:wdp}
b^* \;=\; \arg\max_{b}
      \sum_{i=1}^{K} v^{*}_{\tilde{\theta}_i}(b_i).
\end{equation}

\paragraph{Integer-Programming Formulation}
\begin{subequations}
\label{eqs:ilp}
\begin{align}
\text{Maximise}\quad & \sum_{i=1}^{K}\sum_{j=1}^{m_i} v_{i}(b_{j})\,x_{i,j}
  \label{eq:ilp-objective}\\[4pt]
\text{s.t.}\quad
  & \sum_{j=1}^{m_i} x_{i,j} \;\le 1
    &&\forall i \in \{1,\dots,K\}
    \label{eq:ilp-one-person-one-bid}\\
  & \sum_{i=1}^{K}\sum_{j=1}^{m_i} a_{i,j,k}\,x_{i,j} \;\le 1
    &&\forall k \in \{1,\dots,N\}
    \label{eq:ilp-one-item-one-bundle}\\
  & x_{i,j} \in \{0,1\}
    &&\forall i,j.
    \label{eq:ilp-variables}
\end{align}
\end{subequations}

\subsection{Lindahl Pricing Rule}
\label{appendix:lindahl-pricing-rule}

\begin{equation}
\label{eq:lindahl-prices}
\phi_i(b) \;=\; v^{*}_{\tilde{\theta}_i}(b) \;-\; d_i.
\end{equation}

\section{Bundle-Valuation Subroutine}
\label{appendix:bundle-valuation-subroutine}

This subroutine takes as input the description of a person's preferences, $\mathcal{D}(p)$, and the description of a bundle $b$, $\mathcal{D}(b)$, and gives the value the person has for the bundle, by issuing a LLM call that reasons over the description of the person's preferences and the description of bundle's contents, according to the following pipeline prompt :

\subsection{Pipeline Prompt}
\label{appendix:bundle-valuation-pipeline-prompt}

\begin{MyVerbatim}
Here is the scenario description: (*@\D{S}@*).

Here is the description of a person's preferences who is interested in making a bid in this auction: (*@\D{p}@*).

---

They have the option to receive the following PROPOSED_BUNDLE of items:
(*@\D{b}@*)

---

Please give what you think the person would value this PROPOSED_BUNDLE of items at.

Please always use the following five-step process whenever estimating a person's value for a given bundle given their preferences:
1. Check if the person has explicitly stated the value for that bundle, give this if so
2. Find the closest bundle(s) from the given bundle that the person has explicitly valued
3. Identify the process by which the person has specified they will calculate their value
4. Identify any other relevant criteria
5. Factor (2), (3), and (4) in to estimate the person's value

Give the final value of the bundle of the {bundle.total_quantity()} items in the following format: `Bundle value: \$[value]`.
\end{MyVerbatim}

\subsection{Example Response}
\label{appendix:bundle-valuation-example-transcript}

Below is an example of the LLM's response in a bundle-valuation subroutine call issued for the ELECTRONICS scenario, for the simulated bidder Cecilia described in the example seed, and for the bundle consisting of an iPad Air (M2) and an Apple Pencil (2nd Generation). The LLM reasons through the situation and then gives what the bidder values the bundle at.

\begin{MyVerbatim}
To estimate Cecilia's value for the proposed bundle of items (which includes the iPad Air 11-inch (IPAD12) and the Apple Pencil (2nd Generation) (APPLEPENCIL2)), let's follow the five-step process outlined:

1. **Check if the person has explicitly stated the value for that bundle**: Cecilia has not explicitly stated a value for the bundle consisting of the iPad Air and the Apple Pencil (2nd Generation).

2. **Find the closest bundle(s) from the given bundle that the person has explicitly valued**: 
   - She has explicitly valued the iPad Air (IPAD12) bundled with the Apple Pencil Pro (APPLEPENCILPRO) at $720.
   - She has also valued the iPad (IPAD9) bundled with the Apple Pencil Pro (APPLEPENCILPRO) at $470.

3. **Identify the process by which the person has specified they will calculate their value**: 
   - Cecilia applies a premium value to combinations of compatible items that enhance her functionality. 
   - Since she highly values the Apple Pencil Pro for her creative work, the inclusion of an Apple Pencil (2nd Generation) will likely have a different perceived value compared to the Pro variant.
   - However, she has not specifically stated any value for the Apple Pencil (2nd Generation) or how it compares to the Pro version in this type of bundle.

4. **Identify any other relevant criteria**: 
   - Cecilia would likely apply a discount for the inclusion of the Apple Pencil (2nd Generation) rather than the Apple Pencil Pro. 
   - The iPad Air (M2) bundle was valued higher when combined with the Pro version of the Pencil, indicating she sees higher value in advanced tools that enrich her creative workflow.

5. **Factor (2), (3), and (4) in to estimate the person's value**: 
   - The iPad Air (IPAD12) has a baseline value of $600.
   - The Apple Pencil (2nd Generation) generally might not add as much value as the Pro version, which Cecilia values at $120; however, since it is still a creative tool, she won't discount it too heavily but will assign it a lower value (let's estimate around $80 instead of the Pro's $120).
   - The value of the proposed bundle can be assessed as follows:
     - iPad Air: $600
     - Apple Pencil (2nd Generation): $80
     - Combined Value: $600 + $80 = $680

Since the Apple Pencil (2nd Generation) is not as substantial an upgrade in her creative workflow compared to the Pro version, we may apply a slight adjustment downward, let's say a 10% discount for it not being the Pro model:
- Final estimated value: $680 - ($680 * 0.10) = $680 - $68 = $612.

In conclusion, the estimated value of the proposed bundle is: 

```Bundle value: \$612```
\end{MyVerbatim}

\section{Example Answers to Natural Language Questions}
\label{appendix:natural-language-questions}

This subroutine takes as input the description of a person's preferences, $\mathcal{D}(p)$, and a question $q$, and gives the answer the person would give for the question, by issuing a LLM call that reasons over the descriptions of the person's preferences and the question asked, according to the following pipeline prompt:

\subsection{Pipeline Prompt}
\label{appendix:natural-language-question-pipeline-prompt}

\begin{MyVerbatim}
You are a helpful assistant.

Here is the scenario description: (*@\D{S}@*).

Here is the description of a person who is interested in making a bid in this auction: (*@\D{p}@*).

---

The person is asked the following question: (*@q@*)

---

Please reason through and then craft how you think the person would respond to the question. Keep the answer concise (only one sentence) and relevant to the question. Provide the answer in the following format: `Answer: "[response]"`.
\end{MyVerbatim}

\subsection{Example Response}
\label{appendix:natural-languague-question-example-transcript}

Below are a question and answer for each of the three scenarios. The answer is given by the question-answering subroutine for a specific seed. The first seed is the example seed under the ELECTRONICS scenario.

\begin{MyVerbatim}
ELECTRONICS scenario (example seed)
Q: What specific types of electronics are you interested in purchasing, and what is your maximum bid for those items?
A: I am primarily interested in purchasing the Apple Pencil Pro, with a maximum bid of $120, and I would also consider bundles that include an iPad if they enhance my digital art experience.

PRESERVES scenario
Q: What types of gourmet preserves do you prefer (e.g., fruit, savory, unique flavors), and what is your maximum bid amount for a single jar or a bundle of jars?
A: I prefer health-conscious fruit preserves with low sugar options, specifically the Sugar-Free Raspberry Spread and Organic Strawberry Jam, and my maximum bid for a single jar is $10, while for a bundle including both, I would consider $17.

TRANSPORTATION scenario
Q: How much would you be willing to spend on all three scooters and all three bikes, and do you prefer one over the other?
A: I am only interested in the Schwin Suburban and the Titan Escape 3, so I would be willing to spend around $850 total, but I do not consider the scooters at all.
\end{MyVerbatim}

\section{Equivalence Query Subroutine}
\label{appendix:equivalence-query-subroutine}

The equivalence query response function, $\mathcal{M}^p_E$, is implemented by an LLM call. We prompt the LLM to give the simulated person's response to the equivalence query given a description of the scenario $\mathcal{D}(\mathcal{S})$ and the person's preference $\mathcal{D}(p)$, the hypothesis XOR valuation defined in terms of the atomic bundles $b_i$ and their corresponding valuations $v_i$, and the threshold $\epsilon$. When using the equivalence query, we maintain a memory of prior value query responses during the preference elicitation process and prompt the LLM with it to simplify its task of responding to equivalence queries. Below is the pipeline prompt for the equivalence query LLM call.

\subsection{Standard Equivalence Pipeline Prompt}
\label{appendix:equivalence-query-pipeline-prompt}

\begin{MyVerbatim}
Here is the scenario description: (*@\D{S}@*)

Here is the description of the person's preferences. The person is interested in making a bid in this auction: (*@\D{p}@*)

An XOR preference $\theta = (B, v)$ consists of a set of atomic bundles $B \subseteq \mathcal{B}$ and valuation function $v: B \rightarrow \mathbb{R}_+$. For any bundle $b \in \mathcal{B}$, this preference $\theta$ induces the valuation

$v^*(b) = \max_{\{b' \in B \mid b' \subseteq b\}} v(b')$

This formulation captures the idea that the value of a bundle is determined by the highest-valued atomic bundle it contains. Similarly, prices $\phi$ induce prices $\phi^\*$, where the price of a bundle is the highest price of a bundle it contains. The induced valuations and prices capture the notion of *free disposal*, i.e., the price or valuation of a bundle do not increase when an item in the bundle is removed.

Here the hypothesis XOR valuation function is given as a list of atomic bids:

[[[[ Atomic bid 1 - Bundle: (*@\D{b_1}@*) ; Valued at (*@$v_1$@*) ]]]]
...
[[[[ Atomic bid n - Bundle: (*@\D{b_n}@*) ; Valued at (*@$v_n$@*) ]]]]

Please always use the following five-step process whenever estimating a person's value for a given bundle given their preferences:
1. Check if the person has explicitly stated the value for that bundle, give this if so
2. Find the closest bundle(s) from the given bundle that the person has explicitly valued
3. Identify the process by which the person has specified they will calculate their value
4. Identify any other relevant criteria
5. Factor (2), (3), and (4) in to estimate the person's value

The person has explicitly valued the following bundles:

[[[[ Atomic bid 1 - Bundle: (*@\D{b_1}@*) ; Valued at (*@$v_1$@*) ]]]]
...
[[[[ Atomic bid n - Bundle: (*@\D{b_n}@*) ; Valued at (*@$v_n$@*) ]]]]

Please help me identify which bundles, if any, where the hypothesis XOR valuation function would be the most incorrect in relation to the person's own valuation from their explicit valuation of the bundles and secondarily the description of the person's preferences. Ignore those that are less than (*@$epsilon$@*) off. Ask these key questions to make sure we don't miss anything:

1. What items have not been mentioned that the person would have a valuation over? What is the implicit valuation of items not mentioned explicitly by the person?
2. What bundles of items does the person have specific interest in?
3. Would the person accept modifications to those bundles? If so, what?
4. What about larger bundles that have all or nearly all of the items, or multiple sub-bundles of interest?
5. Choose a random bundle (flip a coin for each item) and see if it should be included

This is call number (*@$num\_calls$@*) of an iterative process to elicit their full XOR valuation. Bundles can have one or many items. A bundle can have all or nearly all of the items. Please give the item types codes and quantities to represent a bundle. Give me five bundles if NOT EQUIVALENT if possible. DO NOT GIVE BUNDLES ALREADY IN THE XOR VALUATION. Valid item type codes: (*@$code_1, \dots, code_n$@*)
\end{MyVerbatim}

\section{Proxy \texttt{value\_query} Prompt}
\label{appendix:proxy-value-query-prompt}

As part of the CECA loop, the proxy must evaluate whether its person is satisfied with a given bundle at posted prices. To do so, the proxy issues a value query by leveraging the conversation history it has accumulated—both from direct natural language interaction and from previous value or demand queries. The following prompt is used to instruct the proxy LLM to estimate the person's valuation for a specific bundle:

\begin{MyVerbatim}
You are helping to represent a person in an auction where we are bidding on their
behalf.

Here is the scenario description:
(*@\D{S}@*)

Here is the current conversation history with the person:
(*@$history\ (includes\ value\ and\ demand\ queries)$@*)

Here is the primary conversation history:
(*@$primary\_history\ (includes\ only\ natural\ language\ queries)$@*)

Please help me infer what you think the person would value the following bundle at:
(*@\D{b}@*)
\end{MyVerbatim}

\section{Proxy \texttt{get\_next\_question} Prompt}
\label{appendix:proxy-get-next-question-prompt}

We now turn to the design of our LLM-based proxy, which is responsible for eliciting a person's preferences through strategic natural language queries. At each iteration, the proxy must decide what to ask next based on its existing conversation history. The prompt below guides the proxy in generating its next question, balancing exploration and informativeness to improve the quality of the inferred XOR bid.

\begin{MyVerbatim}
You are helping to represent a person in an auction where we are bidding on their
behalf.

Here is the scenario description:
(*@\D{S}@*)

Here is the current conversation history with the person:
(*@$history\ (includes\ value\ and\ demand\ queries)$@*)

Here is the primary conversation history:
(*@$primary\_history\ (includes\ only\ natural\ language\ queries)$@*)

What should the proxy ask the person next to better understand their preferences?
Please make sure to get to dollar values, be strategic about what items or groups
of items you ask about to maximise information. If you have a good idea of the
bidder's valuation in general, you can ask about specific bundles of items to
better understand their preferences. Otherwise, ask a general question.
Reason step by step.

Give their next question -- only one -- in the following format:
Question: "[question here]"
\end{MyVerbatim}

\section{Proxy Target/Check/Happy Decision Prompt}
\label{appendix:proxy-target-check-happy-decision-prompt}

Once the proxy has collected sufficient information through value queries and natural language conversations, it must decide how to proceed within the auction. Specifically, it must determine whether to propose a new bundle to target, verify preferences via a demand query (CHECK), or conclude that the bidder is satisfied (HAPPY). This decision is informed by the proxy's internal belief state, price and allocation information, and the transcript of prior interactions with the person.

The following two-part prompt drives this decision-making process. In Part 1, the proxy reasons through the current auction state and conversation history to assess the bidder's likely next move. In Part 2, the proxy must format its output as one of three actions—TARGET\_BUNDLE, CHECK, or HAPPY—with constraints to avoid redundant or invalid proposals.

\subsection{Part 1}
\begin{MyVerbatim}
You are helping to represent a person in an auction where we are bidding on their
behalf.

Here is the scenario description:
(*@\D{S}@*)

Here are the currently tracked bundles:
(*@\D{b_1} \dots \D{b_n}@*)

Here are the current prices:
(*@$p_1\dots p_n$@*)

Here is their currently allocated bundle:
(*@\D{b^*}@*)

Here is your current conversation history:
(*@$history$@*)

Please help me identify a single bundle that the bidder might want to target to
bid on. If there is no such additional bundle, just say that the bidder is
happy. If you want to double-check with the bidder to see what they want, just
say that too.

First, take stock of how the bidder feels about their currently allocated bundle
at current prices. If they have no currently allocated bundle or if there are no
current prices because it is just the start of the auction, **be AGGRESSIVE in
trying out target bundles.**

At other times, prioritise: Test bundles with high value to quantity ratio..

The current target bundles will be shown in the conversation history: make sure
you target at least 3 to 5 bundles before saying the
bidder is HAPPY.

You may or may not choose to CHECK their demand at the given prices. It's a good
idea to do so if you feel like you are missing something.

Place a low priority on giving a `CHECK` response, a
low priority on giving a `TARGET_BUNDLE` response,
and a high priority on giving a `HAPPY` response.

You are a helpful assistant, so without giving a final response give your
thinking below.
\end{MyVerbatim}

\subsection{Part 2}

\begin{MyVerbatim}
Now format to give the appropriate action of either `TARGET_BUNDLE`, `CHECK`, or `HAPPY`. Do not target a bundle already targeted. Here are the items and their quantities available. **Do not give a bundle exceeding the available quantities.**

(*@$items\ and\ quantities$@*)
\end{MyVerbatim}

\end{document}

%% file: figures/auction-structure.tikz
\begin{tikzpicture}[xscale=4, yscale=1.1]
\tikzstyle{description}=[
    rounded corners=3px,
    fill=black!7,
    inner sep=6,
    minimum width=60px,
    font=\normalsize,
]
\tikzstyle{person}=[
    scale=1.7,
    text=orange,
    fill=black!7,
    rounded corners=5px,
    outer sep=3px,
    inner sep=3px,
]
\tikzstyle{proxy}=[
    scale=1.7,
    text=teal,
    fill=black!7,
    rounded corners=5px,
    outer sep=3px,
    inner sep=3px,
]
\tikzstyle{auction}=[
    scale=3.5,
    text=blue,
    fill=black!7,
    rounded corners=5px,
    outer sep=3px,
    inner sep=3px,
]
\tikzstyle{communication}=[
    Latex-Latex,
    draw=black!70,
    ultra thick,
    fill=black!70,
]

\node[person] (person1) at (1,1) {\faUser};
\node[person] (person2) at (1,2) {\faUser};
\node[person] (person3) at (1,3) {\faUser};

\node[proxy] (proxy1) at (2,1) {\faMicrochip};
\node[proxy] (proxy2) at (2,2) {\faMicrochip};
\node[proxy] (proxy3) at (2,3) {\faMicrochip};

\node[auction] (auction) at (3.1,2) {\faChartBar[regular]};

\draw[communication, dashed] (person1) edge (proxy1);
\draw[communication, dashed] (person2) edge (proxy2);
\draw[communication, dashed] (person3) edge (proxy3);

\draw[communication] (proxy1) edge[bend right] (auction);
\draw[communication] (proxy2) edge (auction);
\draw[communication] (proxy3) edge[bend left] (auction);

\node[description, fill=orange!30] at (1,0.1) {People};
\node[description, fill=teal!30] at (2,0.1) {Proxies};
\node[description, fill=blue!30] at (3.1,0.1) {Auctioneer};

\end{tikzpicture}